\documentclass[12pt]{article}

\usepackage{amsmath, amssymb, amsthm}
\usepackage{mathrsfs}
\usepackage{bm}

\usepackage[utf8]{inputenc}
\usepackage[T1]{fontenc}
\usepackage{lmodern}
\usepackage{microtype}
\usepackage{setspace}
\usepackage{hyperref}
\usepackage{graphicx}
\usepackage{epigraph}
\usepackage{float}

\title{\vspace{-2cm}The Information Geometry of UMAP}

\author{Alexander Kolpakov \\ University of Neuch\^atel, Switzerland \\ \href{mailto:kolpakov.alexander@gmail.com}{kolpakov.alexander@gmail.com} 
   \and Aidan Rocke \\ Solomonoff Consulting, Netherlands \\ \href{mailto:rockeaidan@gmail.com}{rockeaidan@gmail.com} }

\date{\today}

\usepackage[a4paper, margin=2.5cm]{geometry}

\usepackage{charter}         

\begin{document}

\maketitle

\begin{abstract}
In this note we highlight some connections of UMAP to the basic principles of Information Geometry. Originally, UMAP was derived from Category Theory observations. However, we posit that it also has a natural geometric interpretation. 
\end{abstract}

\section{Introduction}

The Uniform Mapping and Projection (UMAP) algorithm was first described in \cite{mcinnes2020umap}. This is one of the most efficient clustering algorithms that can be used for unsupervised classification and unsupervised feature extraction among other purposes. The survey article \cite{umap-google} gives a good visual presentation of UMAP's capabilities.  

The derivation of UMAP in \cite{mcinnes2020umap} is heavily based in category theory: our aim is to show that UMAP's fundamental assumptions and techniques have a very natural interpretation via Information Geometry. 

The general setting of UMAP is that we are given a dataset $X = \{X_i\}$ that is sampled from a compact oriented Riemannian manifold $(M, g)$ with boundary in $\mathbb{R}^m$. Let $d: M \times M \rightarrow \mathbb{R}_{\geq 0}$ be the geodesic path metric on $M$. UMAP seeks to embed $X$ into a lower--dimensional space $\mathbb{R}^n$, with $n \ll m$, as a set $Y = \{ y_i \} \subset \mathbb{R}^n$ such that the higher--dimensional proximity between points is preserved in and, moreover, visually revealed if $n = 2$ or $3$. 

Once an embedding $Y \subset \mathbb{R}^n$ is obtained, one may consider $n$ to be the actual dimension (or close to the actual dimension) of $X \subset \mathbb{R}^m$. If $X$ approximates $M$ in any reasonable sense, we may also think of UMAP embedding $M$ in $\mathbb{R}^n$ instead of $\mathbb{R}^m$, with $n \ll m$.  

The following points provide a very basic description of what UMAP does. 

\paragraph{Conformal rescaling.} The $kNN$--graph is created on the points of $X$. Between each pair of high--dimensional data points $X_i$ and $X_j$ the edge probability (or the edge weight, or the degree of belief) is defined as 
$$ p_{i|j} = \exp\left(- \frac{d(X_i, X_j) - \rho_i}{\sigma_i} \right), $$
where $\rho_i$ is the distance from $X_i$ to its nearest neighbour in $X$, and $\sigma_i$ is the rescaling factor. The probabilities are computed for each existing edge $e = i \sim j$ of the $kNN$--graph on $X$.  

The rescaling factors are chosen so that for each $X_i$ we have
$$
\sum_{i: i \sim j} p_{i|j} = \log_2 k. 
$$

\paragraph{High--dimensional probabilities.} For each edge $e=(i,j)$ of the $kNN$--graph in $\mathbb{R}^m$, the symmetrised (high--dimensional) weight $w_h(e) = p_{ij}$ is defined as
$$
p_{ij} = p_{i|j} + p_{j|i} - p_{i|j} \cdot p_{j|i}. 
$$

The symmetrisation process is necessary as the previous step does  \textit{not} guarantee that $p_{ij}$ coincides with $p_{ji}$. 

\paragraph{Low--dimensional probabilities.} So far each edge $e=(i,j)$ in the high--dimensional $kNN$--graph on $X = \{X_i\}$ is assigned a weight $w_h(e)$. For each edge $e = (i,j)$ in the low--dimensional embedding $Y = \{y_i\}$ we will also assign a weight $w_l(e)$. The weight $w_l(e)$ will be determined by the yet unknown $y_i$'s in accordance with the fact that the amount of information conveyed by $X$ should be as close as possible to the amount of information conveyed by $Y$. UMAP assigns the following  weight to each low--dimensional edge $e=(i,j)$ between $y_i$ and $y_j$:
$$
w_l(e) = (1 + a \|y_i - y_j\|^{2b}_2)^{-1},
$$
for some appropriate parameters $a, b > 0$.

\paragraph{Cross-entropy minimisation.} Finally, the cross--entropy of $X$ and $Y$ is minimised. Namely, we write
$$H(X, Y) = - \sum_e w_h(e) \log w_l(e) + (1 - w_h(e)) \log (1 - w_l(e)),$$
and then solve
$$ H(X, Y) \longrightarrow \min_{Y = \{y_i\}}.$$

The minimisation runs over the low--dimensional points $Y=\{y_i\}$, and once an optimal solution is found we obtain the UMAP embedding of $X$. 

\paragraph{Algorithm implementation.} The actual algorithm implementation is available at \cite{umap-github}. The most important issue is that the implemented algorithm is different from the one that is claimed theoretically in \cite{mcinnes2020umap}. As described in \cite{umap-true} due to the sampling strategy that the actual implementation uses for stochastic gradient descent, the loss function turns out to be different from the classical cross--entropy. In our study, however, we concentrate only on theoretical aspects and possible generalisations of underlying ideas. 

\section{The Information Geometry of UMAP}

\subsection{Uniformity assumption}

If we assume the data $X$ to be uniformly distributed on a Riemannian manifold $M$ with metric tensor $g$ having volume forms $\omega$ then, away from the boundary $\partial M$, any ball $B_i$ of fixed volume centered at $X_i$ should contain the same number of points of $X$ regardless of where on $M$ it is centred.

Given finite data and small enough local neighbourhoods, this approximation may be accurate enough even for data samples near $\partial M$.

Conversely, the ball $B_i$ centred at $X_i$ that contains exactly $k$ nearest neighbours of $X_i$ should have approximately the same volume regardless of the choice of $X_i \in X$.

In essence, by creating a custom distance in the neighbourhood of each $X_i$ we can ensure the validity of the assumption of uniform distribution on the manifold by rescaling each ball $B_i$, though the challenge remains to patch the different $B_i$'s together. 

If the balls $B_i$ do not have major overlaps, we may assume that the rescaling is close to conformal, i.e. that it preserves angles, as inside each ball a simple homothety is performed. 

This is what is done in UMAP \cite{mcinnes2020umap} while the higher--dimensional edge weights are computed: we would like to single out this step because of its importance in the understanding and functioning of the entire algorithm. 

In Riemannian Geometry this conformal rescaling technique is related to the classical Moser's trick \cite{moser}. 

Let $\omega_1$ and $\omega_2$ be two volume form on a compact oriented Riemannian manifold $M$ with boundary, such that 
$$\int_M \omega_1(\mathbf{x})\, \mathrm{d}\mathbf{x} = \int_M \omega_2(\mathbf{x})\, \mathrm{d}\mathbf{x}.$$
Then there exists a diffeomorphism $\phi: M \rightarrow M$ such that $\phi^* \omega_2 = \omega_1$. 

By setting 
$$\omega_1(\mathbf{x}) = \frac{\int_M \omega(\mathbf{x}) \, \mathrm{d}\mathbf{x}}{\int_M \mathbf{1}\, \mathrm{d}\mathbf{x}} = \textrm{const};\; \omega_2(\mathbf{x}) = \omega(\mathbf{x}),$$
we get that there exists a volume--preserving diffeomorphism  $\phi: M \rightarrow M$ such that $\phi(M)$ admits a volume form with constant density. Replacing $M$ with $\phi(M)$ having $\phi^*g$ as metric tensor and $\phi^*\omega$ as volume form allows us to perform uniform sampling easily. 

However, the diffeomorphism $\phi$ does not have to be conformal: e.g. the Weyl tensor of $M$ does not have to be preserved by $\phi$. Same applies to the discretized version of the diffeomorphism produced from a ball covering $B_i$ of $M$: approximating $\phi$ inside each ball with $D \phi$ will be a (discrete analogue) of conformal rescaling only if $D \phi$ is a diagonal matrix. 

Moreover, even if the points of $X$ approximate $M$ (e.g. in the Hausdorff distance), it is not certain that their distribution is uniform within each ball $B_i$, and even more so on $M$ itself. The failure of the uniformity assumption will not preclude UMAP from constructing an embedding. However, such an embedding $Y$ may loose most of the features of $M$ that could have been captured by a uniformly sampled $X$. 

\begin{figure}[h]
    \centering
    \begin{minipage}{0.47\columnwidth}
        \centering
        \includegraphics[scale=0.60]{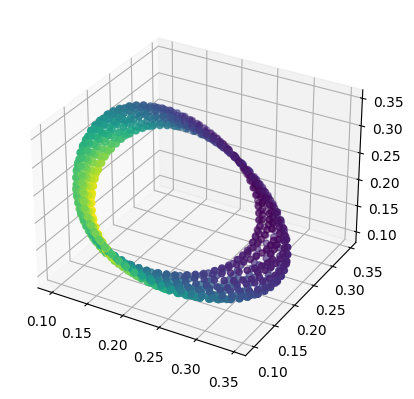}
    \end{minipage}
    ~
    \begin{minipage}{0.47\columnwidth}
        \centering
        \includegraphics[scale=0.60]{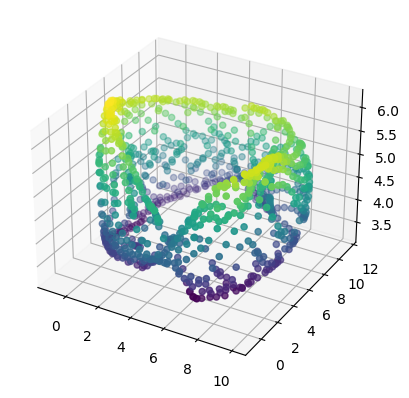}
    \end{minipage}
\caption{The uniform point distribution that follows from the Veselov--Shabat KdV equation: both the $3$D projection (left) and the UMAP embedding (right) resemble $2$D surfaces.}\label{fig:KdV}
\end{figure}

For example, solving numerically the discrete Veselov--Shabat KdV equation produces a uniform point distribution on a $5$--dimensional manifold \cite{pg}, that projects to a $2$--dimensional torus in the $3$--space. In this case, the UMAP embedding looks arguably close to a $2$--dimensional torus in the $3$--space, too, see Fig.~\ref{fig:KdV}. Here we would say that the dimension of the projection and the UMAP embedding coincide. 

\begin{figure}[h]
    \centering
    \begin{minipage}{0.47\columnwidth}
        \centering
        \includegraphics[scale=0.60]{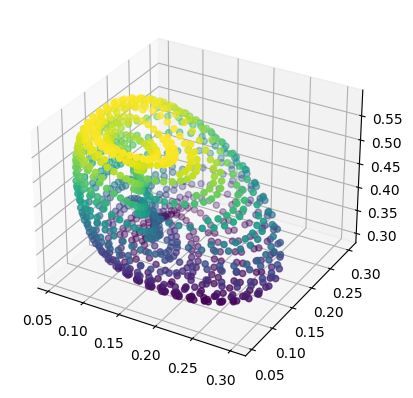}
    \end{minipage}
    ~
    \begin{minipage}{0.47\columnwidth}
        \centering
        \includegraphics[scale=0.60]{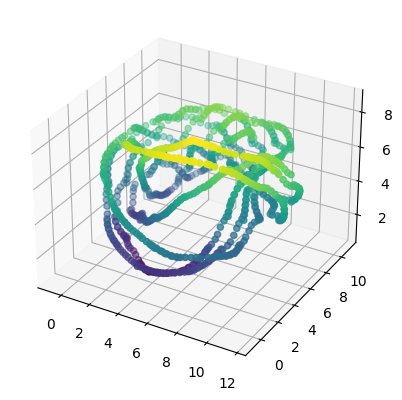}
    \end{minipage}
\caption{A non--uniform distribution: the projection still looks like a $2$D surface, but the UMAP embedding is essentially $1$D.}\label{fig:PG}
\end{figure}

However for the Paik--Griffin replicators described in \cite{pg} the point distribution is not necessarily uniform, although the point clouds produce by them still approximate a $5$--dimensional manifold. This manifold is projected into a $3$--dimensional torus, as before, though the corresponding UMAP embedding looks like a curve. Here, despite the projection looking like a surface, the UMAP embedding is rather $1$--dimensional, see Fig.~\ref{fig:PG}. 

In both computations above (see Fig. \ref{fig:KdV} -- \ref{fig:PG}) the same number of points (1'000) and comparable sets of UMAP parameters were used. Our GitHub repository \cite{github} contains more information about it.

The issue can be mitigated by sampling more points: the only remaining problem is that in this case we had to use a sample that is bigger by an order of magnitude (with 10'000 points instead of 1'000). This is, however, an unsurprising solution: by adding more points to $X$ approximating a manifold we will eventually obtain a distribution in each point's neighbourhood that is close enough to uniform. Indeed, the density of any continuous probability distribution can be approximated by a locally constant function (i.e. a step function).  


\subsection{High--dimensional probabilities}

\subsubsection{Probabilistic $kNN$--graphs}

The high--dimensional probabilities serve the purpose of  recovering the weighted nearest--neighbour graph as an approximation of the dataset's local geometry. To be precise, in the high--dimensional space UMAP uses ``fuzzy'' membership functions (also called ``membership strength'' functions) \cite{fuzzy}
$$
p_{i|j} = \exp\left( - \frac{d(X_i, X_j) - \rho_i}{\sigma_i} \right),
$$
where $d(X_i, X_j)$ is the geodesic distance between $X_i$ and $X_j$ on $M$, $\rho_i$ is the distance from $X_i$ to its nearest neighbour, and $\sigma_i$ is the conformal rescaling factor.

We may consider the $kNN$--graph on $X$ as a random graph where each pair of vertices $X_i$ and $X_j$ is connected according to the Bernoulli random random $\xi_{ij}$, such that for each edge $e=(i,j)$ we have $\xi_{ij} \sim \mathrm{Bernoulli}(p_{i|j})$, and if there is no edge between $i$ and $j$ in the initial $kNN$--graph we simply put $\xi_{ij} \sim \mathrm{Bernoulli}(0)$. 

Then a vertex $X_i$ can pass 
$$
I = \sum_{i: i \sim j} \xi_{ij}
$$
bits of information to its neighbours in the binarised random version of the $kNN$--graph defined above. On the other hand, from the local picture, informing $k$ neighbours would require $\log_2 k$ bits only. Thus, in order to pass information in the $kNN$--graph parsimoniously, we need to keep
$$
I = \log_2 k + O(1).
$$

In order to derandomise, we take the expected value of both sides, and obtain
$$
\mathbb{E}[I] = \sum_{i: i \sim j} p_{i|j} = \log_2 k + O(1),
$$
which is the rescaling normalisation condition of UMAP. 

UMAP uses the following symmetrisation of high--dimensional probabilities:
$$
p_{ij} = p_{i|j} + p_{j|i} - p_{i|j} \cdot p_{j|i}.
$$
Indeed, from the definition we have that $p_{ij} = p_{ji}$. 

Observe that the symmetrisation procedure simply applies the ``OR'' Boolean operator to the local probabilities $p_{i|j}$ and $p_{j|i}$ as if they were independent. 

Symmetrisation is necessary since UMAP needs to adjust the rescaled metrics on $B_i$'s: the degree of belief of the edge $i \sim j$ may not be equal to the degree of belief of $j \sim i$. 

Finally, for an edge $e=(i,j)$ we have that its high--dimensional weight $w_h(e)$ is defined as
$$
w_h(e) = p_{ij}.
$$

\subsubsection{Probability kernels}

The original ``fuzzy'' membership function used to define high--dimensional probabilities in UMAP seems to be just one of many possible probability kernels that might be used depending on the circumstances. We have performed a few numeric experiments that show that using different types of kernels leads to slightly different though similar results. However, the test dataset that we used are standard sandbox Iris \cite{iris}, MNIST \cite{mnist} and Fashion MNIST \cite{fmnist}. 

In each case the clusters were predicted by using HDBSCAN \cite{hdbscan}. As a measure of goodness for clustering we used the adjusted Rand score (ARS) and adjusted mutual information score (AMIS), together with the more geometric silhouette score.

The complete code of our experiments is available on GitHub \cite{github}. Below we reproduce the most important findings. 

First, we choose a variety of functions that may represent the high--dimensional probability kernels. Let $x = \max \{d(X_i, X_j) - \rho_i, 0\}$ and $y = \sigma_i$. Then we may define
$$
p_{i|j} = \left\{  \begin{array}{cc}
    1, & \text{if } x=0 \text{ or } y=0,  \\
     \kappa(x, y) & \text{otherwise;}
\end{array} \right.
$$
where $\kappa(x, y)$ is one of the following probability kernels whose definitions are motivated largely by the author's curiosity. 

Namely, in our experiments we used
\begin{itemize}
    \item the original membership strength kernel $\kappa(x,y) = \exp\left(-\frac{x}{y}\right)$,
    \item the Gaussian kernel $\kappa(x,y) = \exp\left(-\frac{x^2}{2 y^2}\right)$,
    \item the quadratic kernel $\kappa(x,y) = 1 - \frac{x^2}{x^2 + y^2}$,
    \item the Morse kernel $\kappa(x,y) = y - y \cdot (1 - \exp(-x))^2$,
    \item the harmonic oscillator kernel $\kappa(x,y) = 1 + \frac{y}{2}\cdot x^2$,
    \item the constant function $\kappa(x,y) \equiv 1$.
\end{itemize}

Each of the kernels, except Morse's, satisfies $\kappa(0, y) = 1$, and some also satisfy $\kappa(x, y) \approx 0$ for $x \gg 1$, $y > 0$, but not all of them. The Morse kernel satisfies $\kappa(0,y)=y$, while the harmonic kernel is unbounded for $x\gg 1$. 

We performed UMAP with the above kernels on three standard datasets: Iris, MNIST hand-written digits, and Fashion MNIST. The clustering scores are presented in Tables~\ref{tab:iris}, \ref{tab:mnist} and \ref{tab:fmnist}. All comparative illustrations of UMAP embeddings can be found in the GitHub repository \cite{github}. 
~

\begin{table}[H]
\centering
\begin{tabular}{l|lll}
Kernel     & ARS     & AMIS    & Silhouette \\ \hline \\
Membership & 0.45306 & 0.64825 & 0.81799    \\
Gaussian   & 0.45306 & 0.64825 & 0.79939    \\
Quadratic  & 0.56812 & 0.73158 & 0.88512    \\
Morse      & 0.56812 & 0.73158 & 0.91462    \\
Harmonic   & 0.56812 & 0.73158 & 0.94338    \\
Constant   & 0.56812 & 0.73158 & 0.95540    \\          
\end{tabular}
\caption{Different UMAP kernels and their clustering scores for the Iris dataset}\label{tab:iris}
\end{table}
~
~
\begin{table}[H]
\centering
\begin{tabular}{l|lll}
Kernel     & ARS     & AMIS    & Silhouette \\ \hline \\
Membership & 0.93025 & 0.91622 & 0.64939    \\
Gaussian   & 0.92664 & 0.91097 & 0.65802    \\
Quadratic  & 0.93017 & 0.91519 & 0.66941    \\
Morse      & 0.92877 & 0.91388 & 0.67089    \\
Harmonic   & 0.82310 & 0.87173 & 0.52832    \\
Constant   & 0.92090 & 0.90545 & 0.66253    \\          
\end{tabular}
\caption{Different UMAP kernels and their clustering scores for MNIST}\label{tab:mnist}
\end{table}
~
~
\begin{table}[H]
\centering
\begin{tabular}{l|lll}
Kernel     & ARS     & AMIS    & Silhouette \\ \hline \\
Membership & 0.41240 & 0.64771 & 0.57189    \\
Gaussian   & 0.39793 & 0.64229 & 0.61560    \\
Quadratic  & 0.39673 & 0.64099 & 0.63677    \\
Morse      & 0.41967 & 0.64767 & 0.61484    \\
Harmonic   & 0.28016 & 0.59761 & 0.30513    \\
Constant   & 0.39538 & 0.63634 & 0.64772    \\          
\end{tabular}
\caption{Different UMAP kernels and their clustering metrics for Fashion MNIST}\label{tab:fmnist}
\end{table}

From the above tables, one may observe that the constant kernel performs on par (regarding ARS and AMIS) with the membership strength and Gaussian kernels for MNIST and Fashion MNIST. The former outperforms the latter for the Iris dataset. Also, just the simple constant kernel gives the best silhouette score for Iris and Fashion MNIST, while for MNIST the best silhouette is given by the Morse kernel. As the reader may notice, there are other differences between various kernels, some significant and some relatively minor, for the same set of parameters. 

Thus, the high--dimensional $kNN$--graph does not become entirely binarised (edge or no edge) by the UMAP algorithm, at least not the way it is claimed in \cite{umap-true}. 

\subsection{Low--dimensional probabilities}

Let $Y = \{ y_i \}$ be the image in $\mathbb{R}^n$ of $X$ under UMAP. For the low--dimensional space with ambient dimension $n \leq 3$ so that dataset visualization tools are applicable, the UMAP authors use the following approximation to the Student's $t$--distribution:
$$
w_l(e) = (1 + a \|y_i - y_j\|^{2b}_2)^{-1},
$$
where $e = (i,j)$ is the edge between $y_i$ and $y_j$ in the low--dimensional graph.

As observed in \cite{mcinnes2020umap}, the low--dimensional weight of $e=(i,j)$ is approximately $\sim 1.0$ when $\| y_i - y_j\| \leq d_{min}$ and $\sim 0.0$ otherwise, when the hyperparameters are  $a=1.93$, $b=0.79$ and $d_{min}=0.001$.

We may note that the Student's $t$--distribution is a sensible choice when the variance of the distribution is unknown.

\subsection{On the equivalence of cross--entropy and KL--divergence}

The Kullback--Leibler divergence is widely used in Information Geometry as a distance function which is, however, not symmetric and thus cannot represent a metric. 

Another measure of difference of one distribution from the other is their cross--entropy, which is widely used in practical tasks.

Let the set of all possible edges in the $kNN$--graph be $E$, where we have Bernoulli probability $w_h(e)$ for the edge $e$ in the high--dimensional space and Bernoulli probability $w_l(e)$ in the low--dimensional space.

Then the Kullback--Leibler divergence of the graphs associated with $X$ and $Y$ equals
\begin{equation*}
D_{KL}(X || Y) = \sum_{e \in E} w_h(e) \log \left(\frac{w_h(e)}{w_l(e)}\right) + \sum_{e \in
E} (1-w_h(e)) \log \left(\frac{1-w_h(e)}{1-w_l(e)}\right) = 
\end{equation*}

\begin{equation*}
    = -\sum_{e \in E} w_h(e) \log w_l(e) - \sum_{e \in E} (1-w_h(e)) \log (1-w_l(e)) 
\end{equation*}
\begin{equation*}
    +\sum_{e \in E} w_h(e) \log w_h(e) + \sum_{e \in E} (1-w_h(e)) \log (1-w_h(e)) = 
\end{equation*}

\begin{equation*}
    = H(X, Y) - H(X),
\end{equation*}

where $H(X, Y)$ is the cross--entropy of the high--dimensional and low--dimensional graphs on $X$ and $Y$, respectively, and $H(X)$ is the entropy of the high--dimensional graph on $X$. 

Since the optimisation necessary for the low--dimensional embedding begins once the high--dimensional probabilities are already defined, we have that $H(X) = \text{const}$, and thus
\begin{equation*}
    D_{KL}(X || Y) = H(X, Y) + \text{const}. 
\end{equation*}

This implies that the KL--divergence and the cross--entropy loss functions induce the same learning dynamics for lower--dimensional similarities. 

The above analysis, however, including optimization of the cross--entropy $H(X, Y)$ (also mentioned in \cite{mcinnes2020umap}) eludes the following problem. The low--dimensional set $Y$ may be initialised in a number of ways, including spectral embedding, or PCA, or just randomly. The fact that the points of $Y$ corresponding to connected vertices of the high--dimensional $kNN$--graph on $X$ move into positions minimizing the above cross--entropy does not imply that the points of $Y$ corresponding to the points of $X$ being far apart will also be comparably far apart in the low--dimensional embedding. In other words, not only we need to insure that near in $X$ is near in $Y$, but also that far in $X$ is reasonably far in $Y$.  

This requires that we introduce some sort of repulsion force between the points of $Y$ that correspond to non--connected vertices in the $kNN$--graph of $X$. From the information--geometric prospective, this amounts to adding the Kullback--Leibler divergence for the complement of said $kNN$--graph. 

As we remarked above, a missing edge $e\notin E$ corresponds to $w_h(e) = 0$, thus the additional ``repulsive'' component of the Kullback--Leibler divergence is
$$
D_{KL}(\overline{X}||Y) = - \sum_{e\notin E} \log(1 - w_l(e)) + \sum_{e\notin E} \log(1 - w_h(e)) = - \sum_{e\notin E} \log(1 - w_l(e)) + \mathrm{const}.
$$

Now, we have to minimise the linear combination 
$$
\mathcal{L}(X, Y) = D_{KL}(X||Y) + \alpha \cdot D_{KL}(\overline{X}||Y) = H(X, Y) + \alpha \cdot H(\overline{X}, Y) + \mathrm{const},
$$
where $\alpha > 0$ is the repulsion coefficient. Depending on the sampling mode of the actual computational routine, $\alpha$ can be either a parameter, or can be chosen with respect to certain intrinsic conditions (see \cite[Section 7.3]{umap-true} for a related discussion). 

\subsection{Future research: Vietoris--Rips complexes}

Instead of the high--dimensional $kNN$--graph on $X$, we may build the Vietoris--Rips complex (the $VR$--complex, for short) on $X$. Let $\delta > 0$ be a positive constant. Then the $VR$--complex on $X$ is the abstract simplicial complex with vertices $\{X_i\}$ and any subset $X_{i_0, \ldots, i_d} = \{ X_{i_0}, \ldots, X_{i_d} \}$ forms a $d$--dimensional simplex whenever $X_{i_0, \ldots, i_d}$ can be placed inside a ball in $\mathbb{R}^n$ of diameter at most $\delta$. 

Varying $\delta$ changes the ``zoom'' of the $VR$--complex, and allows to capture essential topological features at different levels of coarseness. The birth and death of topological features is described by the persistence homology \cite{persistence} of $X$, and the latter can likely be used to determine the threshold values of $\delta$ for which the lower--dimensional embeddings of $X$ should be studied. 

As before, the starting point is to compute the probabilities $p_{i|j}$ for each $1$--simplex in the $VR$--complex on $X$. However, in this case we posit that the degree of belief should be the same for all edges of any given simplex. The function 
$$
\wedge(x, y) = x + y - x\cdot y
$$
is associative, as follows from a direct check. It is also mapping the unit square $I^2 = [0,1]^2$ exactly onto $I = [0,1]$. 

Thus we may define the high--dimensional weight of each edge of $X_{i_0, \ldots, i_d}$ inductively by setting
\begin{align*}
    \textrm{Initialize } p \leftarrow p_{i_0|i_1}; \\
    \textrm{Do } p \leftarrow \wedge(p, p_{i_k|i_l}) \textrm{ for each remaining edge } (i_k, i_l); \\ 
    \textrm{Output } \wedge(X_{i_0, \ldots, i_d}) \leftarrow p.
\end{align*}

Then for each edge $e$ of $\Delta = X_{i_0, \ldots, i_d}$ we simply put $$w_h(e, \Delta) = \wedge(X_{i_0, \ldots, i_d}),$$
which defines the high--dimensional weights of the entire $VR$--complex on $X$. 

If an edge $e$ of $X$ belongs to several simplices $\Delta_1, \ldots, \Delta_k$, which is often the case, we can average out the contributions of each of the simplices by setting
$$w_h(e) = \Pi_{i=1}^k w_h(e, \Delta_i)^{1/k}.$$

The rest of the algorithm proceeds without change, and the low--dimensional weights are computed with respect to $1$--simplices (edges) only. 

This approach may allow capturing more essential topology of the high--dimensional data, though at the cost of building a richer combinatorial structure of the $VR$--complex. Indeed, the $VR$--complex can be thought of as a multigraph or, more conveniently for our purposes, as the clique complex of the following $\delta$--neighbours graph. 

Let $G_\delta(X)$ be the graph where vertices are the points of $X$, and whenever two points $u, v \in X$ are at a distance $\leq \delta$, then $u$ and $v$ are connected in $G_\delta(X)$ by an edge. Then the $VR$--complex of $X$ at scale $\delta > 0$ can be though of as the clique complex of $G_\delta(X)$. As the number of cliques in a graph can be exponentially large in the number of vertices (which does not apply to the number of edges, that is at most quadratic), the problem is computationally hard. 

Nevertheless, certain approaches can handle $VR$--complexes with $\sim 20$ millions of simplices in about $80$ seconds \cite{bm, r, zam}, and thus an attempt at a practical generalization of UMAP seems realistic.   

\section{Conclusions}

In this note we produce a theoretical discussion of several geometric aspects of UMAP as compared to its original ``fuzzy'' or ``categorical'' interpretation. We posit that Information Geometry can be the right framework for understating UMAP. 

For example, some of its assumptions have a clear geometric interpretation (uniformity and conformal rescaling) and explanation (normalisation condition), and the purported loss function is nothing else but the $KL$--divergence. 

Thus, the optimisation step of UMAP seeks to embed the high--dimensional probability distribution on the $kNN$--graph of $X$ into a lower--dimensional parameter space $Y$ with the least possible distortion. Then the Fisher metric defined by high--dimensional probabilities is brought as close as possible to the Fisher metric defined by low--dimensional probabilities. If $X$ approximates $M \subset \mathbb{R}^m$, then $Y$ approximates a low--dimensional isometric embedding of $M$ into $\mathbb{R}^n$, with $n \ll m$. 

We also suggest a way to take the topology of Vietoris--Rips complexes into account: these complexes capture the essential topology of high--dimensional data at different levels of coarseness, and embedding them instead of $kNN$--graphs may help discovering hidden structures in the original data. This relates directly to persistence homology and opens up a way of using topological data analysis to produce meaningful embeddings.

\section{Data availability}

All data and computer code are accessible on GitHub \cite{github} and referenced within the manuscript.

\section{Acknowledgements}

A.K. was supported by the Swiss National Science Foundation project PP00P2--202667. A.A.R. thanks the University of Neuch\^atel for hospitality during his research stay in July 2023. Both authors are grateful to the anonymous referee for useful comments and suggestions.

\end{document}